# Decoding Neural Activity to Assess Individual Latent State in Ecologically Valid Contexts

Stephen M. Gordon[1], Jonathan R. McDaniel[1], Kevin W. King[1], Vernon J. Lawhern[2], Jonathan Touryan[2]

[1] *DCS Corporation, Alexandria, VA*

[2] *DEVCOM Army Research Laboratory, Aberdeen Proving Ground, MD*

## ABSTRACT

Currently, there exists very few ways to isolate cognitive processes, historically defined via highly controlled laboratory studies, in more ecologically valid contexts. Specifically, it remains unclear as to what extent associated patterns of neural activity observed under such constraints actually manifest outside the laboratory in a manner that can be used to make an accurate inference about the latent state, associated cognitive process, or proximal behavior of the individual. Improving our understanding of when and how specific patterns of neural activity manifest in ecologically valid scenarios would provide validation for laboratory-based approaches that study similar neural phenomena in isolation and provide meaningful insight into the latent states that occur during complex tasks. We argue that domain generalization methods, borrowed from the work of the brain-computer interface community, have the potential to capture high-dimensional patterns of neural activity in a way that can be reliably applied across experimental datasets in order to address this specific challenge. We previously used such an approach to decode phasic neural responses associated with visual target discrimination. Here, we extend that work to more tonic phenomena such as internal latent states. We use data from two highly controlled laboratory paradigms to train two separate domain-generalized models. We apply the trained models to an ecologically valid paradigm in which participants performed multiple, concurrent driving-related tasks while perched atop a six-degree-of-freedom ride-motion simulator. Using the pretrained models, we derive estimates of the underlying latent state and associated patterns of neural activity. Importantly, as the patterns of neural activity, assessed via the latent state model, become more similar to those patterns observed in the training data, we find changes in behavior and task performance that are consistent with the observations from the original, laboratory-based paradigms. We argue that these results lend ecological validity to the original, highly controlled, experimental designs and provide a methodology for understanding the relationship between observed neural activity and behavior during complex tasks.

# 1.0 INTRODUCTION

Neuroscientific research often requires highly controlled environments with well-defined task parameters in order to isolate brain activity associated with an endogenous state or exogenous stimuli. Employing imaging methods such as electroencephalogram (EEG), scientists obtain high-dimensional measurements of brain activity that can be precisely aligned with experimental conditions. To obtain meaningful and statistically robust associations between experimental conditions and brain activity, data from repeated observations are often aggregated using methods such as ensemble averaging of evoked responses. Based on the assumption of a linear model with an additive noise term independent of experimental conditions, such aggregation requires consistency at the trial or block level. For example, when studying the neural correlates of visual perception, researchers typically measure brain activity while simultaneously presenting a stream of visual stimuli at discrete intervals in rapid succession (Duncan-Johnson & Donchin, 1982; Polich, 1998). In such experiments, the participant may be instructed to detect a particular target stimulus (e.g., "watch for a red circle"), or the stimuli themselves may have intrinsically arousing properties (e.g., a spider). Similar analytical approaches have been used to assess the neural dynamics of slowly varying latent states with experimental designs that induce more tonic responses. Examples of this can be found in the works of (Brouwer, et al., 2017; Ries, et al., 2016) where the authors investigated the changes in neural activity associated with increased mental load during visual search.

While the use of controlled, repeatable paradigms coupled with ensemble analysis approaches have uncovered fundamental properties of brain activity, these methods inherently divorce the brain and its function from its more natural environment, that is, the real world (McDowell, et al., 2013). Despite growing attention to this issue, it remains an open question to what extent associated patterns of neural activity, observed under such constraints, actually manifest outside the laboratory in a manner that can be used to make an accurate inference about latent state and current or future behavior (van Atteveldt, et al., 2018; Parsons, 2015; Sonkusare, et al., 2019). Addressing this question remains a challenge for several reasons. The first is the inherent complexity and high dimensionality of the neural activity itself, even during simple tasks. EEG-based laboratory studies of brain activity sample data from an underlying multidimensional feature space that typically includes scalp topography (space) and spectral bands (frequency). Even low-dimensional EEG recording systems can include hundreds of distinct dependent variables.

Second, natural environments are uncontrolled and real-world tasks are dynamic and multifaceted. In the laboratory, participant data is sampled along a specific 'axis of variability' induced by the independent variables of the experimental paradigm. While a particular axis may be unique to an experimental paradigm, movement along this axis, measured through both behavioral and neural activity, is believed to reflect a more fundamental underlying latent state or cognitive process. Therefore, the experimental design is often intended to maximize across-condition variability while minimizing variability from other exogenous or endogenous sources (Figure 1, left), thus isolating the associated latent states or processes. In ecologically valid domains, such tight control is not possible (Figure 1, right), and a specific cognitive process may become obscured by concurrent variance and corresponding brain activity induced by other aspects of the task or environment.

Third, while computational models, such as those commonly employed by the Brain-Computer Interface (BCI) community, can be useful tools, highly parameterized participant-specific methods run the risk of introducing additional confounds by overfitting to the new participant or task in question. In other words, BCI models that are fit to the task or participant in question make it difficult to retroactively validate that the original patterns of neural activity observed in the laboratory are representative of the activity recorded in the new context. Furthermore, these methods are often dependent on large amounts of training data for both the task and participant in question. However, BCI models that do not require participant- or task-specific data may be useful for decoding neural activity in complex domains.

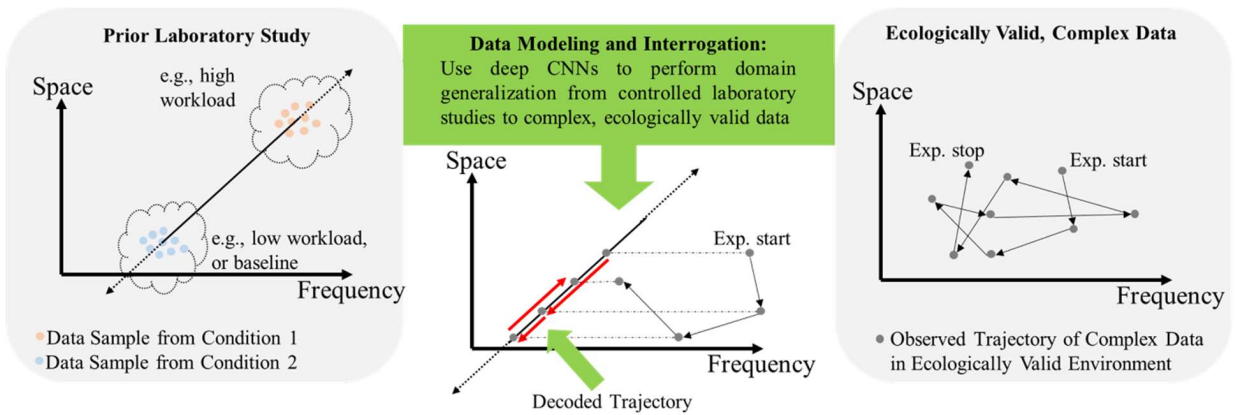

*Figure 1. Deep convolutional neural networks can be used to bridge the gap between a) controlled laboratory studies (left), in which conditions are designed to isolate single latent states and tasks are often low dimensional, and b) ecologically valid domains, in which tasks are multifaceted and neural activity may reflect movement of more than one latent state.*

The purpose of the current work is to show how deep modeling and domain generalization methods adapted from the BCI community can be used to assess the extent to which specific neural activity, originally observed in highly controlled laboratory environments, manifests in complex, ecologically valid environments. To this end, we map latent states, originally observed in those laboratory paradigms, to a small number of brain activity dimensions that can then be translated across tasks and contexts using deep convolutional neural networks (CNNs). The trained CNNs are then used to decode the complex, ecologically valid data from the perspective of the prior, controlled laboratory studies (Figure 1, center).

## 2.0 METHODS AND DATA

Broadly speaking, the setup of our problem can be formulated as content-based retrieval (CBR), sometimes known as query-by-content, which formulates queries based on the content of a data item rather than its labeling or metadata and has been previously proposed for EEG analysis (Su & Robbins, 2013). In CBR tasks, a machine learning model is trained using a given sample of data and tasked with retrieving new samples from a larger corpus of data, and then ranking those samples by their similarity ("content") to the original input pattern, or query. In the current work, the input patterns are those neural patterns observed in the different experimental conditions of the laboratory-based training data (e.g., Condition 1 or 2 in Figure 1, left). The test data are data from an ecologically valid domain. The machine learning task is to rank the instances from the test data as being more similar to the patterns observed in Condition 1 or Condition 2 of the training data. Subsequent analyses then investigate whether *movement* along this axis (i.e., between Condition 1 and Condition 2) is meaningful in the test data. As such, we take special care during the validation stages to show that the domain generalization approach can accurately capture the neural patterns of interest and retrieve similar instances of those patterns in a more complex domain.

### *2.1. MODEL SELECTION*

There have been many prior approaches to using machine learning tools to perform dimensional reduction of EEG data for subsequent analysis. This includes within-participant approaches to analyze both phasic, single trial, responses (Gerson, et al., 2006; Philiastides, et al., 2006; Salazar-Gomez, et al., 2017; Ehrlich & Cheng, 2019; Lopes-Dias, et al., 2019) and tonic latent state measurements (Szafir & Mutlu, 2012; Ehrlich, et al., 2014). There has also been work across participants but within experiments using both shallow and deep methods (Stock, 2016; Vahid, et al., 2018; Vahid, et al., 2020; Banville, et al., 2021). In each of these cases, the goal was to train a machine learning model to detect specific patterns of neural activity in otherwise difficult to analyze conditions, or to decode the EEG signal into a feature space that highlights the differences that can occur across conditions. However, these approaches either require participant- or paradigm-specific calibration or are specifically developed for a single analysis domain (e.g., sleep scoring).

To alleviate the need for participant- or paradigm-specific training data, one can use domain generalization methods (Gordon, et al., 2017). Such methods learn a representation of neural activity in one domain (e.g., participant set and/or task set), and must apply the learned representation to decode data from a new domain (i.e., participant and/or task set). The majority of this research has been conducted by the BCI community, specifically in areas now described as passive BCI (Zander & Kothe, 2011). Domain generalization methods have previously been used to model states related to vigilance (Ma, et al., 2019; Kim, et al., 2022), detect emotions (Li, et al., 2022; Liang, et al., 2022), diagnose neurological conditions (Ayodele, et al., 2020), as well as perform more classical BCI tasks, such as motor imagery (Han, 2021). These methods tend to use deep networks as they have been shown to be effective at modeling data from multiple individuals.

To perform the CBR and neural decoding required in the current work, we argue that it is important to select a model that is known to be accurate in transfer tasks and that is well calibrated, meaning that the outputs should vary smoothly as a function of the similarity of the test data to one or more of the training patterns. Previous work has shown that modern neural networks, including CNNs, possess these qualities, as long as they are framed as binary discrimination tasks (Niculescu-Mizil & Caruana, 2005; Guo et al 2017).

To better explain why it is important that model outputs vary smoothly, we use the work of (McDaniel, et al, 2018) as an example case. In that work, a domain-generalized CNN was applied across experiments to detect target versus nontarget visual stimuli in a rapid visual presentation paradigm. The authors used the EEGNet model of (Lawhern, et al., 2018) and showed that the outputs of the CNN were more sensitive to nontarget distractors that were conceptually similar to the target than the outputs of other models trained solely within participants. In other words, the outputs of the CNN changed smoothly for nontargets as a function of the similarity of the nontarget images to the target category. This is in line with prior neuroscientific research that has repeatedly shown that distractors often elicit weakened, but target-like, neural responses (Azizian, et al., 2006; Watson, et al. 2005; Sawaki, et al., 2006; Marathe, et al., 2015) that degrade smoothly as a function of target-nontarget similarity.

Follow-on work in (Solon, et al., 2019) used the same CNN model and a collection of four different experiments that each involved a variation of the visual-evoked target detection paradigm. The authors showed that such a domain-generalized deep CNN could i) faithfully reproduce the underlying signal statistics, for example, variations in the amplitude and latency of the neural response, and ii) improve the overall SNR. The authors concluded by using the trained model to interrogate a novel and complex dataset in which participants performed free-viewing visual searches and found that the CNN-decoded signal could help dissociate visually cued vs. fixation-locked responses.

Finally, (McDaniel, et al., 2020) used the same domain generalization approach to decode the tonic latent state during a driving task. In that work, the authors used a publicly available Nap EEG study (Mei, et al., 2018) as the training set and a simulated nighttime driving task as the target domain. The authors found that as individual latent states *drifted* towards the "asleep" end of the spectrum, the rate of driving errors increased significantly. The work of (Solon, et al., 2019) and (McDaniel, et al., 2020) used CNNs trained for binary discrimination tasks to decode the changes in neural activity in a test dataset along the principal axis of the original training set. However, each of these prior works was limited to a single interrogative model applied to a single, highly controlled test dataset. In other words, while establishing a proof of concept, it did not demonstrate the generalization of the method to complex domains. In this work, we use the EEGNet architecture (Lawhern, et al., 2018) to instantiate the two CNN models as it meets the criteria previously set forth (i.e., transfer accuracy and designed for binary tasks); however, we argue that any domain-generalized neural network model that has been validated for accuracy and whose outputs change smoothly as a function of the similarity of the input test patterns to the original training data could be used.

## 2.2. Model Training

The first CNN model was trained using an experiment in which participants were exposed to alternating periods of single-modality (visual) and dual-modality (visual and auditory) tasks. We refer to this dataset as DS_LOAD. A description of this dataset is provided in Section 2.3. We refer to the model trained using this dataset as the LOAD model, because the experimental design modulated task load.

The second CNN model was trained using an experiment designed to investigate alert versus drowsy driving in a simulated night-time setting. In this experiment, which we refer to as DS_ALERT, the participants operated a simulated vehicle along an empty stretch of road at night. Lateral perturbations simulating a crosswind pushed the vehicle out of the lane, which required corrective responses. A description of this dataset is provided in Section 2.3. This model is referred to as the ALERT model.

During model training, we hold out a sample of training data to generate validation loss curves. The final model weights were selected based on the evaluation of the loss curve. Once trained, we further validated both models using hold-out experimental datasets, referred to as DS_VALIDATE_LOAD and DS_VALIDATE_ALERT. Finally, we applied both trained models to a complex test set, referred to as DS_COMPLEX. In DS_COMPLEX, the participants operated a simulated vehicle while perched atop a ride-motion-simulator. Participants were responsible for performing multiple, concurrent driving-related tasks, and in certain conditions, had access to an autopilot that could be used to assist in some aspects of the task. A description of the validation and test datasets is provided in Section 2.3. Table 1 provides an overview of the five experimental datasets used in this work.

*Table 1. Overview and Summary of the Five Experimental Datasets Used*

| DATA SET | DESCRIPTION | SUMMARY |
|---|---|---|
| DS_LOAD (Brouwer, et al., 2017) | **Single vs. Dual Tasking:** <br> 2 conditions: <br> 1. Visual only <br> 2. Visual with auditory math | 17 Participants <br><br> 32-channel BioSemi ActiveTwo |
| DS_ALERT (Lin, et al., 2005) | **Alert vs. Drowsy Driving:** <br> 1 Condition: <br> Drive at night and respond to lateral perturbations of the vehicle | 14 Participants <br><br> 33-channel Neuroscan NuAmps |
| DS_LOAD_VALIDATE (Ries, et al., 2016) | **Single vs. Dual Tasking:** <br> 4 conditions: <br> 1. Visual with auditory ignored <br> 2. Visual with auditory (N back level = 0) <br> 3. Visual with auditory (N back level = 1) <br> Visual with auditory (N back level = 2) | 17 Participants <br><br> 64-channel BioSemi ActiveTwo |
| DS_ALERT_VALIDATE (Mei, et al., 2018) | **Manually Annotated Sleep Study:** <br> 1. Periods of pre-sleep wakefulness <br> 2. Sleep stage 1 <br> 3. Sleep stage 2 | 22 Participants <br><br> 64 channel antiCHamp active electrode system |
| DS_COMPLEX (Metcalfe, et al., 2017) | **Complex Driving Task:** <br> 3 conditions: <br> 1. Manual driving <br> 2. Driving with speed control option <br> 3. Driving with speed & lane control option | 18 Participants <br><br> 64-channel BioSemi ActiveTwo |

2.2.1. <u>Data Preparation</u>

To prepare the data prior to training or testing, we resampled each dataset to 128 Hz and to the following 30 channel montage, based on the 10-20 configuration: Fp1, Fp2, F7, F3, Fz, F4, F8, FT7, FC3, FCz, FC4, FT8, T7, C3, Cz, C4, T8, TP7, CP3, CPz, CP4, TP8, P7, P3, Pz, P4, O1, Oz, and O2. This channel set represents the largest common set of channels across all the datasets. Each of the five datasets that we used was collected at a higher sampling rate than the final 128 Hz selection. Each dataset contained two mastoid signals, which were averaged and used as references, except for DS_ALERT_VALIDATE, which contained only a single mastoid signal that was used as a reference. We bandpass-filtered the data between [0.3, 50] Hz by first low-pass filtering at 50 Hz using a finite-impulse response filter (FIR) and then high-pass filtering at 0.3 Hz using another FIR filter. We then performed median absolute deviation normalization for the data from each participant in each experiment. Other than the channel downselection step, these are the same preprocessing steps described by (Solon, et al., 2019).

*2.2.2.  Data Cleaning*

For each participant from each training dataset, we performed ICA decomposition of the data to identify artifactual eye movement components. For DS_LOAD, which possessed an eye-tracking signal from an external eye tracker, we used the IMICA algorithm (Gordon, et al., 2015) to perform decomposition. This algorithm is a constrained optimization approach to the decomposition problem that uses reference signals to order and improve the quality of the signal separation. Using IMICA, we computed and removed two components: horizontal eye movement and vertical eye movements and blinks. For the DS_ALERT dataset, which did not have simultaneously recorded eye movements or EOG signals, Infomax ICA (Bell & Sejnowski, 1995) was used for decomposition. We visually inspected the data to identify eye movement and blink components. These components were removed whenever they were present. The ICA cleaned data were used to create a 'clean' version of each training dataset, in which eye movement artifacts were removed. We performed this step as a precaution to remove artifacts that were common to all datasets and could potentially become discriminative features for the CNN model.

*2.2.3. Model Training*

The EEGNet architecture (Lawhern, et al., 2018) was used to implement each CNN model. EEGNet is a compact (i.e., low number of free parameters) CNN that was specifically designed for EEG data and has been shown to provide state-of-the-art classification within participants (Lawhern, et al., 2018) for numerous BCI applications. Here, we fit the EEGNet model with eight temporal filters, four spatial filters per temporal filter, and 16 separable filters (EEGNet 8-4-16, using the notation from (Lawhern et. al., 2018)). We used a temporal filter length of 64 samples, representing 0.5 sec of data sampled at 128 Hz. The model was trained for (up to) 150 iterations using the Adam optimizer with default parameter settings (Kingma & Ba, 2014), and a mini-batch size of 16 instances, optimizing a binary cross-entropy loss function. The dropout probability was set to 0.25 for all layers.

To create the complete training data, we first divided the continuous EEG data from each participant into 3 second epochs at 128 Hz with 0.5 second window overlap. We used 3 second epochs for several reasons. First, the initial layers of EEGNet perform the temporal and spatial filtering of the input EEG data. The 3 second window provides *roughly* three cycles of information from the lowest frequency band (i.e., Delta, which starts at approximately 0.5 – 1 Hz). Second, prior empirical testing within the training sets did not reveal significant changes in performance at longer time windows (i.e., 4-5 seconds) but did show small decrements in performance at shorter time windows (i.e., ~2 seconds).

Next, we labeled the partitioned training epochs using the predefined experimental condition, or block, extracted from the experimental design. In the case of class imbalance, we performed randomized downsampling of the majority class per participant. We did not balance across participants within a single experimental dataset. Based on prior experience, this would be too limiting, as some participants in a given experiment may have only a handful of trials. Using the previously stated settings for EEGNet resulted in a model with 2,801 trainable parameters. Using the training datasets described in later sections yielded training sets with sizes of ~16,000 epochs for the LOAD model and ~6,000 epochs for the ALERT model.

After the initial partitioning and balancing of the training data, we concatenated the training data with the same training samples but drawn from the ICA cleaned version (i.e., eye movement artifacts were removed). This was done as a form of data augmentation, which has been shown to improve the generalizability and robustness of machine learning models across a variety of domains (Shorten & Khoshgoftaar, 2019), including BCI decoding (Lotte, 2015; Zhang, et al., 2019). Finally, we used an ensemble approach by training five distinct instantiations of the CNN, using five distinct and unique downsamplings for each training set. Given the number of downsamplings that can occur in the data to ensure class balance, the ensemble approach allows us to use more available data while ensuring class balance. During the testing, we averaged the outputs of these five instances to create a single output signal.

*2.2.4. Model Application*

To interrogate the new data, we convolved the models over the processed EEG signals for one sample at a time (Figure 2). Throughout the remainder of this paper, we refer to the *middle of the three second input window* as the application time point for the model. In other words, the model outputs are produced for a given time point, t, using data from the window: [t-1.5, t+1.5] seconds. As stated in the previous section, we applied an ensemble of trained CNNs and averaged the outputs to obtain a single value. This single value varied over time, as a function of the underlying neural patterns in the test data (Figure 2). To collapse the variations in outputs for the cross-participant analysis, we z-scored the averaged model output time series over the entire set of available data from a given participant.

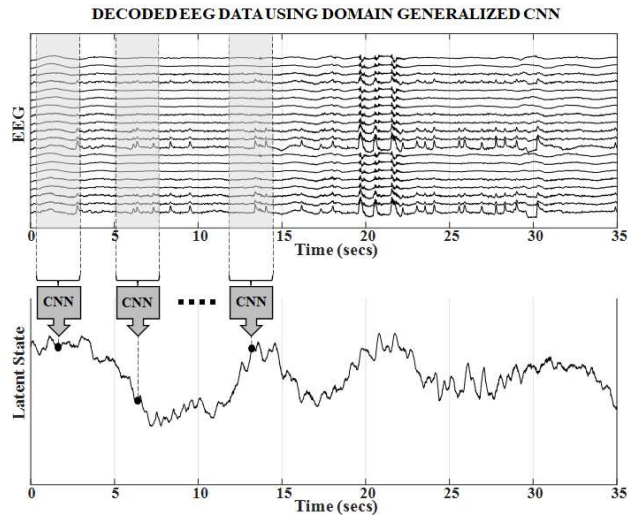

*Figure 2. The trained CNN is convolved over the test data one sample at a time to produce a latent state time series signal.*

## 2.3. DATA DESCRIPTION

In the following sections, we describe the data used to train the LOAD and ALERT models. We also describe the data used for the initial validation and the final test data.

*2.3.1. DS_LOAD*

Participants (N=21, nine males, 12 females) between the ages of 19 and 30 years (average age:23) either performed a guided visual search task and an auditory math task simultaneously (dual-task, high workload) or a guided visual search task only (single-task, low workload) (Brouwer, et al., 2017). The auditory task was presented in both cases to provide consistent auditory stimulation, but participants were instructed to ignore this component in the single-task condition. EEG data from each participant were recorded using a 32-channel BioSemi ActiveTwo system digitized at 512 Hz. Four EOG electrodes were used to record horizontal and vertical eye movements, and a single external electrode was placed on each of the left and right mastoids to provide the reference signals. Data from four participants were excluded based on data quality and/or missing event information (e.g., stimulus presentation times), resulting in 17 participants being included in this study.

The guided visual search task required participants to fixate within a prearranged grid of visual stimuli, with the fixation location dictated by the appearance of a marker. The dual-task component was a binaurally presented auditory math task in which participants were given a sequence of numbers and operations, and instructed to compute the total. Participants performed eight blocks of 11 trials per condition. Participants were instructed to ignore the math task or compute their total score in alternating successive blocks. To create the training data, we selected epochs from the dual-task condition to represent a high workload (high model output) and epochs from the single-task condition to represent a low workload (low model output).

*2.3.2. DS_ALERT*

Participants (N = 14) between the ages of 20 and 35 years underwent a simulated driving experiment, in which each participant had to maintain the lane position while responding to laterally directed perturbations (Lin, et al., 2005a). The perturbations were designed to simulate mild-to-medium crosswinds. EEG data were recorded using a 33-channel Compumedics Neuroscan system digitized at 500 Hz. One channel was used to record the vehicle data and two other channels were used to record the mastoid references. The length of the experiment and nature of the task were designed to induce drowsiness/sleep. On average, the experiment lasted for 60 min.

During post-processing, the reaction times (RT) were computed for each trial (i.e., perturbation). Unlike DS_LOAD, which came with clean experimental conditions, the periods of drowsy driving in the DS_ALERT dataset were best measured using participant behavior. For each participant, the RTs were normalized and used to divide the data into alert and drowsy epochs. This behavioral-based definition was used to ensure consistency with prior work and analysis of the given dataset. We used the top 10% for low alertness (low model output) and the bottom 10% for high alertness (high model output) of the trials to define the classes for training. This dataset has been analyzed and studied several times within the BCI community (Lin, et al., 2005b; Wu, et al., 2016; Cui & Wu, 2017; Wu, et al., 2015; Hajinoroozi, et al., 2015). Although it is a simulated nighttime driving study, latent state changes are believed to be related to the onset of drowsiness and sleep.

*2.3.3. DS_VALIDATE_LOAD*

Participants (N=17, all males) with an average age of 32.8 years performed a guided visual search task in parallel with a second binaurally presented auditory task (Ries, et al., 2016). EEG data from each participant were recorded using a 64-channel BioSemi ActiveTwo system digitized at 512 Hz. Four external electrodes were used to record bipolar horizontal and vertical EOG signals, and a single external electrode was placed on each of the left and right mastoids to provide the reference signals. Fourteen participants were included in the original study, with three additional participants later added, resulting in 17 participants.

The visual search task for this experiment required participants to follow a red annulus around the screen and press a button if the annulus stopped at a prespecified target. The auditory task for this experiment was an N-back matching task in which participants listened to a string of numbers presented at approximately 2 second intervals. In the baseline condition, participants were presented with both visual and auditory stimuli, but were instructed to ignore the auditory component. Next, were three dual-task conditions with N-back levels of N=0, N=1, and N=2.

*2.3.4. DS_VALIDATE_ALERT*

Participants (N=22, 16 male) with an average age of 25.5 years were asked to take brief naps after performing a visual working memory task (Mei, et al., 2018). EEG data from each participant were recorded using a 64 channel antiCHamp active electrode system (Brain Products, GmbH). A single mastoid reference was used during the data recording. In post-processing, the data were manually scored to identify periods of pre-sleep wakefulness (Awake) and sleep stages 1 and 2 (Stage 1, Stage 2). For this work, we discarded the visual working memory component of the task.

*2.3.5. DS_COMPLEX*

Participants (N=18) performed a simulated driving task (Figure 2, left), while perched atop a six-degree-of-freedom ride motion simulator (Metcalfe, et al., 2017). EEG data from each participant were recorded using a 64-channel BioSemi ActiveTwo system. Four external electrodes were used to record bipolar horizontal and vertical EOG signals, and a single external electrode was placed on each of the left and right mastoids to provide the reference signals. Eighteen participants were included in the original study, but one was excluded because of missing data.

In the task, participants were instructed to maintain the lane position, maintain a consistent distance to a lead vehicle that was ahead of their own, maintain the appropriate speed, and watch for pedestrians. If pedestrians entered the path of the vehicle, they were instructed to push a button on the steering wheel, which would cause the simulated pedestrian to vanish, thereby avoiding a collision. Similar to the work of (Lin, et al., 2005a), a mild-to-moderate crosswind was used to occasionally perturb the vehicle from its lane. In addition, the lead vehicle adjusted its speed at random times, forcing the participant to take similar compensatory measures to maintain the vehicle distance. Speed limit signs were posted along the simulated course to adjust vehicle speed. The participants operated the vehicle along a closed course in both turns and straight ways. Unlike the work of (Lin, et al., 2005a), however, there was no focus on producing drowsiness, and at no point in the experiment did any participant drift off to sleep.

The participants performed these driving tasks under three experimental conditions. The first condition was a Manual Driving mode in which the participant was responsible for all aspects of the driving task. In the remaining conditions, participants could turn on, at their discretion, an autonomy that would take over some of the driving tasks. Under one condition, termed Speed Only, autonomy offered only speed control. In this condition, the autonomy, once enabled, would maintain the vehicle speed and leader-follower distance. In the other condition, termed Full Autonomy, the autonomy was designed to perform both speed and lane control tasks. In either autonomy condition, turning on the system required pressing a toggle switch located on the side of the steering column, similar to turning on cruise control in most modern vehicles. Turning off the autonomy involved simply intervening in the vehicle's controls (i.e., pushing the brake or accelerator or turning the steering wheel).

## 3.0 RESULTS

In this section, we first validate the performance of the domain-generalized models on the two hold-out validation datasets. Each validation dataset was selected to have a conceptual overlap with one of the training sets. The purpose of this validation is to show that 1) the domain-generalized models reliably and correctly respond when presented with data that should match, in some way, the underlying neural patterns on which the model was trained and 2) the model outputs changed predictably and smoothly as the neural features moved from one end of the spectrum to the other.

After this validation step, we then applied each model to DS_COMPLEX, analyzed the model outputs as a function of the experimental condition, and measured human behavior (i.e., turning on/off the autonomy). We analyzed human performance as a function of the model output, that is, the predicted latent state.

## 3.1. VALIDATION RESULTS

Figure 3, A shows the performance of the LOAD model when applied to DS_VALIDATE_LOAD. Here we see that, as expected, model outputs correctly respond to the latent state changes, and underlying neural pattern changes, that occur between the single-task (Ignore) condition and the high workload, dual-task (N=2) condition. To produce this plot (as well as Figure 3, B), model outputs were z-scored per participant, averaged per condition, and then averaged across participants. Performing an analysis of variation (ANOVA) test on the averaged samples per participant we find statistically significant difference ($p < 0.01$) between the Ignore and N=2 condition. In addition, and just as importantly, the N=0 and N=1 conditions are placed *between* these two ends of the spectrum and showing a steady increase along the spectrum "Ignore" to "N=2". Naturally, changes in the underlying latent state are not expected to be linear with N-back level but are expected to be monotonically increasing.

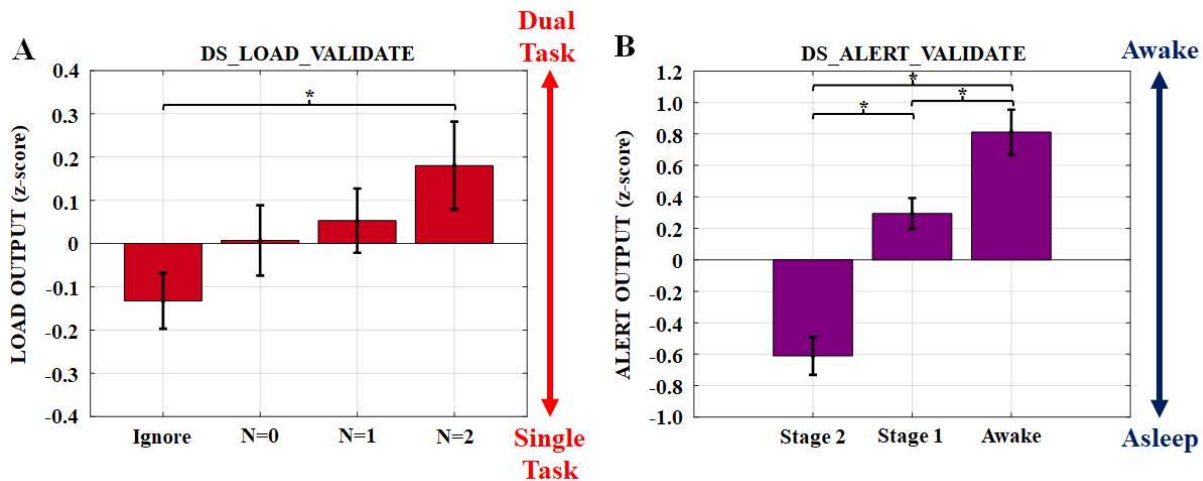

*Figure 3. A) Average LOAD model outputs for the DS_LOAD_VALIDATE dataset (\* indicates model outputs are significantly different at p < 0.01). B) Average ALERT model outputs for the DS_ALERT_VALIDATE dataset, with significant differences indicated by (\*).*

Figure 3, B shows the performance of the ALERT model when applied to DS_ALERT_VALIDATE. Again, we see that, as expected, model outputs correctly respond to the latent state changes in the underlying data. Model outputs show monotonic changes from Stage 2 sleep, to Stage 1 sleep, and then to Awake, with each latent state producing significantly different ($p < 0.01$) mean output levels (again using ANOVA tests across participants).

## 3.2. DS_COMPLEX RESULTS

We applied the LOAD and ALERT models to the DS_COMPLEX data. Figure 4 shows the average model output for both the LOAD and ALERT models for the three conditions in that dataset: Manual Driving, Full Autonomy, and Speed Only autonomy.

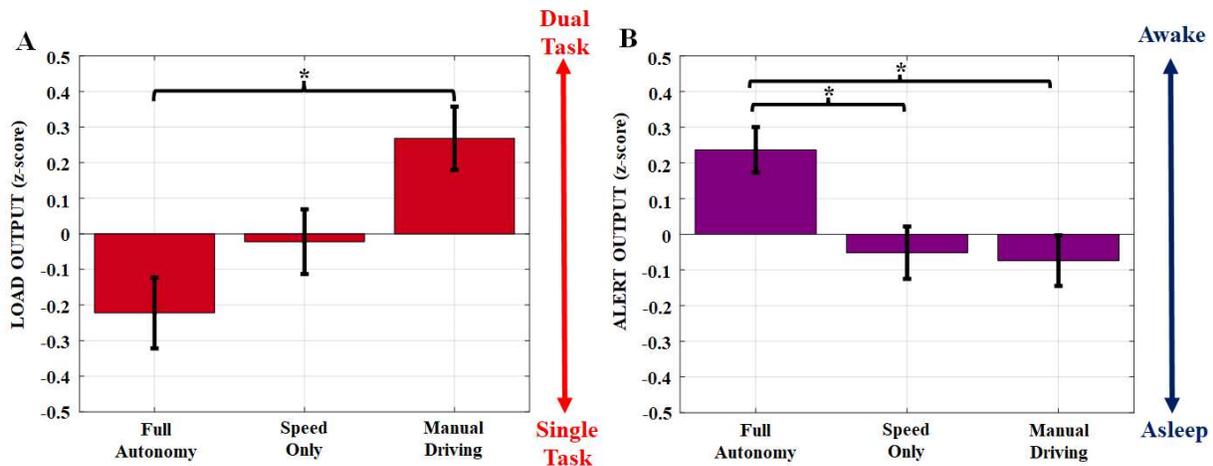

*Figure 4. A) LOAD model outputs for each driving condition in DS_COMPLEX. B) ALERT model outputs for each driving condition in DS_COMPLEX. (\*) indicates significant differences ($p < 0.01$).*

Here, we see that both models indicate a change in the underlying latent state across conditions, with the Full Autonomy condition producing the most distinct latent state. Again, we averaged the model outputs per condition and across the participants. We also performed an ANOVA on the mean model outputs per condition across participants with significant changes ($p < 0.01$), as indicated by (*) in Figure 4. However, while the LOAD model indicates a monotonic shift from Full Autonomy, to Speed Only autonomy, to Manual Driving, the ALERT model groups the Speed Only autonomy and Manual Driving together. In other words, for the Manual Driving condition, the LOAD model indicates that the underlying patterns of neural activity are most similar to those observed in the high-workload condition of DS_LOAD and that this similarity decreases as tasking is removed. The ALERT model considers the Full Autonomy condition to be most similar to the wakeful driving periods in DS_LOAD, whereas Speed Only autonomy and Manual Driving are equivalent. It is beyond the scope of this current work to investigate *why* the ALERT patterns are this way, but we hypothesize that overseeing the autonomy in the Full Autonomy condition does not require much task load, but did require higher levels of vigilance, alertness, or latent states associated with these concepts. It should be noted that because the CNN outputs were first z-scored per participant the scales in Figure 4, and throughout the rest of the section, are relative to DS_COMPLEX only and cannot be used to equate the conditions in DS_COMPLEX with those in either of the validation datasets.

Next, we analyzed model outputs in the autonomy conditions around the events of turning on and off the autonomy. The data were aggregated across the Full Autonomy and Speed Only condition. To account for gross state differences, such as those observed in Figure 4, we further z-scored the data per condition and averaged the CNN model outputs for each participant for the window of time [-25, -5] secs before the switch event and [+5, +25] secs after the switch event. We left a 10 second gap around the switch event to ensure that motor artifacts and other neural processes did not interfere with the analysis. We then averaged the CNN model outputs across participants. We excluded any autonomy switch event that was less than 30 secs from the previous event or the next event. The results are presented in Figure 5, where we see that there is a significant decrease in LOAD model outputs when the autonomy is turned on ($p < 0.01$), and then a significant increase ($p < 0.01$) when the autonomy is turned off (Figure 5, A). There were no significant changes in the ALERT model outputs around these events (Figure 5, B).

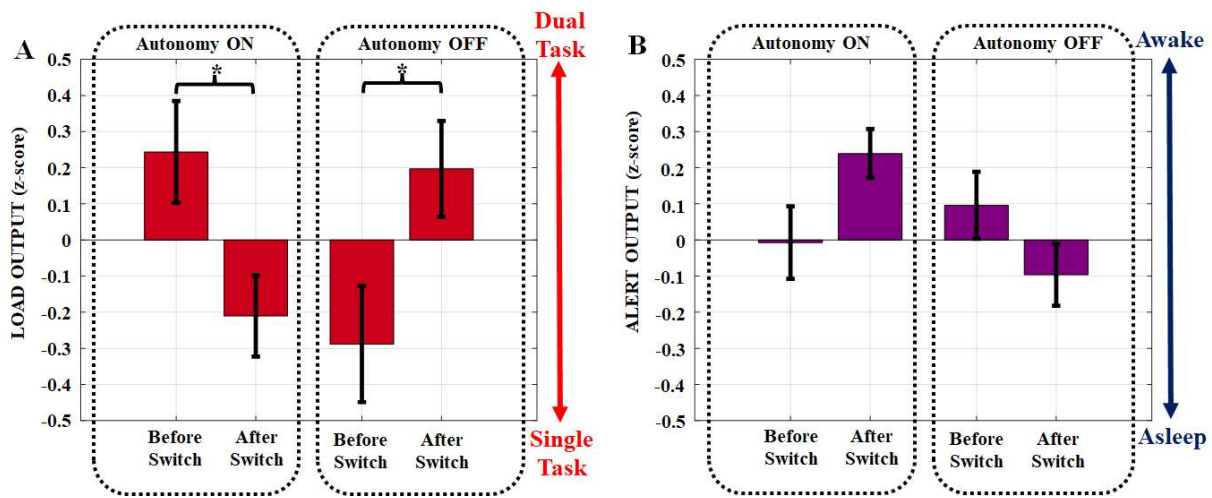

*Figure 5. A) Mean model output around autonomy switch events for the LOAD model. The switch to Autonomy ON is marked by a significant ($p < 0.05$) decrease in LOAD model outputs. The switch to Autonomy OFF is marked by a significant ($p < 0.01$) increase in LOAD model outputs. B) Mean model output around autonomy switch events for the ALERT model. No significant changes in outputs values were found.*

Up to this point, the results have been condition- and behavior-based, with the LOAD model outputs showing significant changes in the neural activity across conditions in which task load changed. To understand how continuous changes in neural activity impact behavior and task performance, we analyzed the Manual Driving condition only as a function of the estimated latent state. More specifically, we investigated the rate of driving errors in the Manual Driving conditions. We computed the probability of observing a given model output value and the probability of committing a driving error, given the output value. In this dataset, driving errors included: 1) drifting out of lane, 2) failure to follow the speed limit, 3) failure to maintain the leader-follower distance, or 4) hitting a pedestrian. The results of this analysis are shown in Figure 6. To contribute to this analysis, each participant had to have at least 10 driving errors. A total of 11 participants met this criterion with an average of 28.7 (±22.4) driving errors with a minimum value of 10 and a maximum value of 82. The probability distributions in Figure 6 A and B represent the average across these 11 participants. The probability distributions shown in Figure 6 A and B are estimates provided to highlight the relationship between the model output and driving performance. To obtain smooth estimates for discussion purposes, we used bin sizes of 0.75 stepped across the data with a step size of 0.1. Here, we show only the results from the ALERT model, as there was no observed relationship between the LOAD model values and the driving errors. Inspecting the results (Figure 6, A), we see that as the ALERT model output values decrease from +2.0 s.d. to -2.0 s.d., the probability of committing a driving error increases by approximately six-fold. Gray bars indicate scale. We then analyzed these data to determine whether there was a statistical relationship between model output and driving performance. These results are shown in Figure 6, C. To do this, we used non-overlapping bins with bin centers at -2, -1, 0, 1, 2, and a bin size of 1. We computed the normalized error rate per person in these bins as the percentage of the total errors that occurred in each bin divided by the total time spent in each bin for each person. We used a linear mixed model to assess the error rate as a function of the model output, with participants as a random effect. Using this approach, we found significance for both the y-intercept and model output values ($p < 0.001$).

Next, we repeated this analysis using the model output values with a time delay of 8 seconds (i.e., *before* the driving error). We did this to assess the stability of the estimates and to remove suspicion of effects from any artifacts that could have occurred at the moment of analysis. We examined the relationship between model outputs leading up to the driving error and performance. The results are shown in Figure 6, B. While the relationship is not as strong as that observed in Figure 6, A, the linear mixed model analysis applied to the data in Figure 6, D still revealed significant effects of model output on the error rate with p < 0.01.

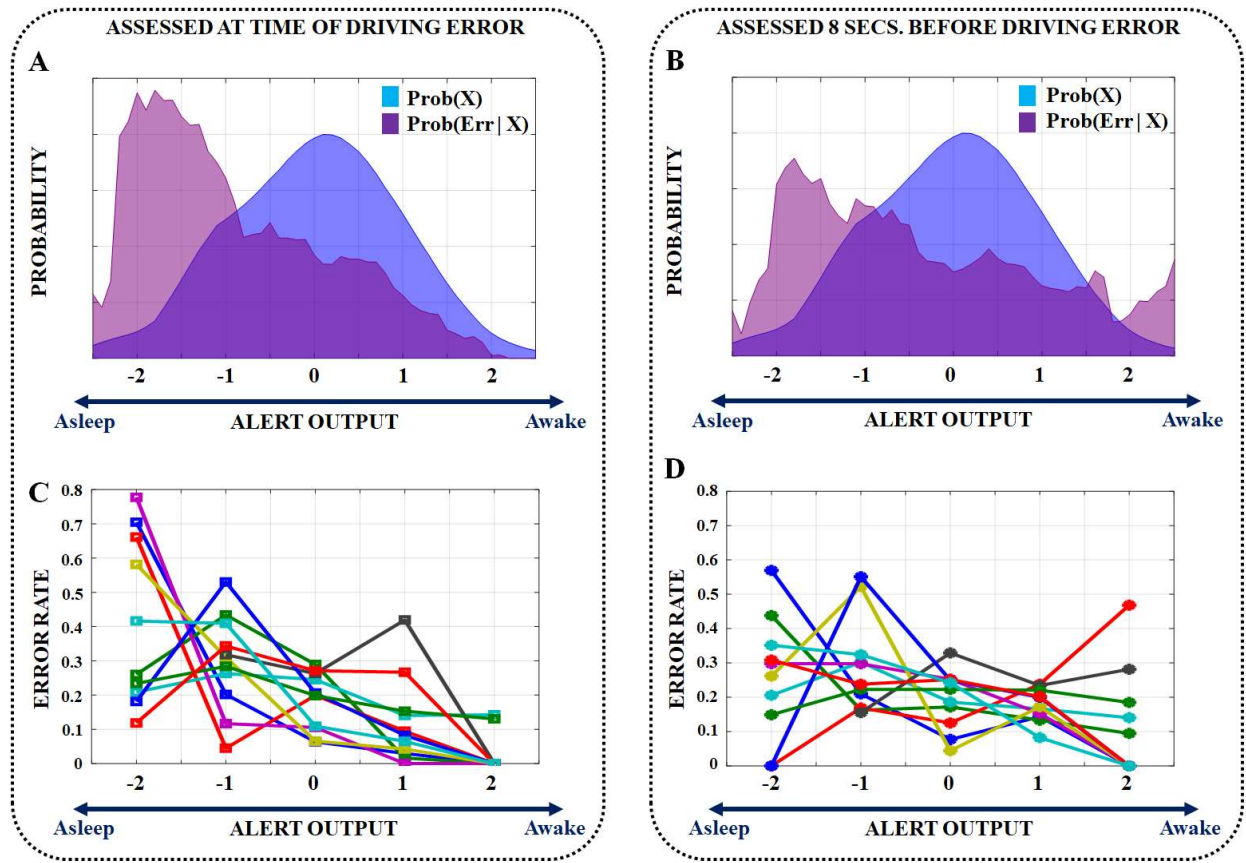

*Figure 6. Probability of committing a driving error on the Manual Driving Condition. A) Probability of observing a given Alert model output value (blue) and the probability of a committing a driving error given that ALERT value assessed at the moment the driving errors occurred (purple). B) Probability of observing a given Alert model output value and the probability of a committing a driving error given that ALERT value assessed 8 secs. before the driving errors occurred. C) Proportion of driving errors committed in each bin, using a bin size of 1 with centers at [-2, -1, 0, 1, 2] for each participant using model output values assessed at the moment of the driving error. D) Proportion of driving errors committed in each bin, using a bin size of 1 with centers at [-2, -1, 0, 1, 2] for each participant using model output values assessed 8 secs. before the driving error.*

## 4.0 DISCUSSION

We first demonstrated that domain-generalized CNN models could reliably detect changes in neural activity in novel validation datasets obtained from well-controlled laboratory studies. We also showed that the CNN outputs changed smoothly as the latent states in the data moved along the axis from one state to another (i.e., from single task to dual task with a high N-back level in DS_VALIDATE_LOAD, or from awake to asleep in DS_VALIDATE_ALERT). In other words, the CNN outputs changed smoothly as a function of the similarity of the neural activity in the test sets to the patterns of neural activity originally encoded in the training sets. Although other deep CNN models may also have this property, it was beyond the scope of the current work to analyze and compare the growing wealth of domain-generalized models for such tasks. Rather, we simply contend that having a model that is both accurate and well-calibrated (i.e., sensitive to changes in the underlying neural activity) is important.

Applying the same models to DS_COMPLEX allowed us to uncover blocks of time in that dataset, in which the neural activity was, again, more (or less) similar to the patterns of neural activity encoded in the distinct, well-controlled conditions of the training datasets. As task load increased in DS_COMPLEX across conditions (i.e., from Full Autonomy to Speed Only autonomy to Manual Driving), there were consistent changes in the outputs of the LOAD model, indicating that the underlying neural activity moved closer to that observed in the dual-task (visual and auditory) condition of the training dataset, DS_LOAD. Similar shifts were also observed when task load abruptly changed due to the participants manually turning on or off the autonomy. It should be noted that the visual/auditory balance of tasking in DS_COMPLEX was not set such that one condition required both visual and auditory tasking, while another condition required visual tasking alone. Given this analysis and approach, therefore, if one were to ask if the patterns of neural activity and the changes between them observed in the well-controlled but artificial tasks of DS_LOAD are representative of patterns (and changes) that could occur in complex domains, without regard to the specific source of task demands (i.e., visual or auditory), it is our perspective that the answer would have to be *yes*.

The same was true for the patterns of neural activity observed in the original DS_ALERT dataset. Using this model, the analysis of DS_COMPLEX revealed that the unrestricted movement of that latent state during the Manual Driving condition was highly predictive of task performance, specifically, the probability of committing a driving error. Furthermore, the relative changes in the latent state mattered, that is, the changes in the latent state were not indicative of a binary state change but rather revealed movement along a spectrum, and this movement was tied to performance. The rate of driving errors increased almost linearly as the underlying patterns of neural activity drifted towards the "asleep" end of the spectrum.

We should point out that unlike the large domain jump between DS_LOAD and DS_COMPLEX, there were domain similarities between DS_ALERT and DS_COMPLEX. In particular, both datasets utilized driving tasks in which alertness or vigilance was believed to be responsible for the performance. However, DS_ALERT used a highly impoverished scenario with an experimental design intended to put the participants to sleep. We contend that this explains why the ALERT model was able to decode

DS_VALIDATE_ALERT (i.e., sleep study) so well. DS_COMPLEX, on the other hand, was much more representative of everyday driving, and no participants went to sleep during the study.

One analysis that we did not perform was to investigate the specific feature representations learned and encoded by the CNN for latent state discrimination. We argue that this analysis may be misleading. The latent states used in this paper may manifest in various ways through complex interdependencies in the EEG signal. It is erroneous to suggest that there is only one definitive representation for a given latent state. Regarding artifacts, we chose highly controlled laboratory studies to model the latent states to minimize the impact of artifacts. We also removed common artifacts, such as eye movements, and used this clean data to augment the CNN during training; however, this does not eliminate the possibility of artifact contamination or contribution, especially if those artifacts co-occurred with a given latent state.

As for the utility of this approach to the BCI and pBCI, communities, we see no reason why the work presented here could not be used to guide method development in the service of state classification or task performance prediction; however, that was not the focus of the article. Many of the methods that we employed were not those commonly used for real-time EEG processing or analysis. As a simple example, we z-scored model outputs to facilitate cross-participant analysis; however, such a step is not valid in many pBCI applications. Furthermore, it was beyond the scope of the current work to provide additional scientific insight into the specific latent states associated with DS_LOAD and DS_ALERT. Rather, the goal was to investigate where these latent states were meaningful beyond the confines of the original, highly controlled laboratory studies in which they were defined and could be used to make inferences about cognitive processing or behavior in a complex, ecologically valid domain? Based on the current analysis, we would state that *movement of latent state along the axes originally defined in DS_LOAD and DS_ALERT, along with the concomitant changes in neural activity, was meaningful with respect to observed behavior and task performance levels in DS_COMPLEX*.

## 5.0 CONCLUSION

We set out to establish a methodological approach for isolating cognitive processes, principally defined by highly controlled laboratory studies, in more ecologically valid contexts. We trained two CNNs using previously collected experimental data from highly controlled laboratory settings with well-isolated patterns of neural activity. We then applied these to both well-controlled and ecologically valid datasets. We believe that the techniques presented here provide a computational pathway to advance neuroscientific research outside the confines of highly controlled, repeatable paradigms. Furthermore, this approach did *not* require participant- or paradigm-specific training data. For neuroscience studies conducted in impoverished, repeatable paradigms, the current work provides a means to demonstrate that those patterns and the associated axes of variability are meaningful beyond their original laboratory confines. For studies conducted in less-controlled contexts, this approach may provide a pathway for partitioning and analyzing such data. There is much future work to be done. The results, especially those from DS_COMPLEX, raise several questions. Some of these include: 1) what are the interactions between the latent states, 2) what are the relative timescales for each latent state, and 3) are there underlying dynamics that can be modelled on the scale of seconds, or even minutes?

## 6.0 ACKNOWLEDGEMENTS

The authors would like to thank Michael Nonte for helpful discussions involving the conceptual development of this work and would like to thank Drs. Amar Marathe and Jason Metcalfe for providing motivation in the form of discussions as well as complex real-world data with challenging analytical problems. This project was sponsored by the US Army Research Laboratory under Cooperative Agreement Number W911NF-10-2-0022. The views and conclusions contained in this document are those of the authors and should not be interpreted as representing the official policies, either expressed or implied, of the US Government. The US Government is authorized to reproduce and distribute reprints for Government purposes, notwithstanding any copyright notation herein.

## 7.0 REFERENCES

Ayodele, K.P., Ikezogwo, W.O., Komolafe, M.A. and Ogunbona, P., 2020. Supervised domain generalization for integration of disparate scalp EEG datasets for automatic epileptic seizure detection. Computers in biology and medicine, 120, p.103757.

Azizian, A., Freitas, A.L., Watson, T.D. and Squires, N.K., 2006. Electrophysiological correlates of categorization: P300 amplitude as index of target similarity. *Biological psychology*, *71*(3), pp.278-288.

Banville, H., Chehab, O., Hyvärinen, A., Engemann, D.A. and Gramfort, A., 2021. Uncovering the structure of clinical EEG signals with self-supervised learning. *Journal of Neural Engineering*, *18*(4), p.046020.

Bell, A.J. and Sejnowski, T.J., 1995. An information-maximization approach to blind separation and blind deconvolution. *Neural computation*, *7*(6), pp.1129-1159.

Brouwer, A.M., Hogervorst, M.A., Oudejans, B., Ries, A.J. and Touryan, J., 2017. EEG and eye tracking signatures of target encoding during structured visual search. *Frontiers in human neuroscience*, *11*, p.264.

Cui, Y. and Wu, D., 2017, November. EEG-based driver drowsiness estimation using convolutional neural networks. In *International Conference on Neural Information Processing* (pp. 822-832). Springer, Cham.

Duncan-Johnson, C.C. and Donchin, E., 1982. The P300 component of the event-related brain potential as an index of information processing. *Biological psychology*, *14*(1-2), pp.1-52.

Ehrlich, S.K. and Cheng, G., 2019. A feasibility study for validating robot actions using eeg-based error-related potentials. *International Journal of Social Robotics*, *11*(2), pp.271-283.

Ehrlich, S., Wykowska, A., Ramirez-Amaro, K. and Cheng, G., 2014, November. When to engage in interaction—And how? EEG-based enhancement of robot's ability to sense social signals in HRI. In *2014 IEEE-RAS International Conference on Humanoid Robots* (pp. 1104-1109). IEEE.

Farwell, L.A. and Donchin, E., 1988. Talking off the top of your head: toward a mental prosthesis utilizing event-related brain potentials. Electroencephalography and clinical Neurophysiology, 70(6), pp.510-523.

Gerson, A.D., Parra, L.C. and Sajda, P., 2006. Cortically coupled computer vision for rapid image search. *IEEE Transactions on neural systems and rehabilitation engineering*, *14*(2), pp.174-179.

Gordon, S.M., Jaswa, M., Solon, A.J. and Lawhern, V.J., 2017, March. Real world BCI: cross-domain learning and practical applications. In Proceedings of the 2017 ACM Workshop on An Application-oriented Approach to BCI out of the laboratory (pp. 25-28).

Gordon, S.M., Lawhern, V., Passaro, A.D. and McDowell, K., 2015. Informed decomposition of electroencephalographic data. *Journal of neuroscience methods*, *256*, pp.41-55.

Guo, C., Pleiss, G., Sun, Y. and Weinberger, K.Q., 2017, July. On calibration of modern neural networks. In International conference on machine learning (pp. 1321-1330). PMLR.

Hajinoroozi, M., Jung, T.P., Lin, C.T. and Huang, Y., 2015, July. Feature extraction with deep belief networks for driver's cognitive states prediction from EEG data. In *2015 IEEE China Summit and International Conference on Signal and Information Processing (ChinaSIP)* (pp. 812-815). IEEE.

Han, D.K. and Jeong, J.H., 2021, February. Domain generalization for session-independent brain-computer interface. In 2021 9th International Winter Conference on Brain-Computer Interface (BCI) (pp. 1-5). IEEE.

Kim, D.Y., Han, D.K., Jeong, J.H. and Lee, S.W., 2022, October. EEG-based Driver Drowsiness Classification via Calibration-Free Framework with Domain Generalization. In 2022 IEEE International Conference on Systems, Man, and Cybernetics (SMC) (pp. 2293-2298). IEEE.

Lawhern, V.J., Solon, A.J., Waytowich, N.R., Gordon, S.M., Hung, C.P. and Lance, B.J., 2018. EEGNet: a compact convolutional neural network for EEG-based brain–computer interfaces. *Journal of neural engineering*, *15*(5), p.056013.

Li, Y., Chen, H., Zhao, J., Zhang, H. and Li, J., 2022. Benchmarking Domain Generalization on EEG-based Emotion Recognition. arXiv preprint arXiv:2204.09016.

Liang, S., Su, L., Fu, Y. and Wu, L., 2022. Multi-source joint domain adaptation for cross-subject and cross-session emotion recognition from electroencephalography. Frontiers in Human Neuroscience.

Lin, C.T., Wu, R.C., Jung, T.P., Liang, S.F. and Huang, T.Y., 2005. Estimating driving performance based on EEG spectrum analysis. *EURASIP Journal on Advances in Signal Processing*, *2005*(19), pp.1-10.

Lin, C.T., Wu, R.C., Liang, S.F., Chao, W.H., Chen, Y.J. and Jung, T.P., 2005. EEG-based drowsiness estimation for safety driving using independent component analysis. *IEEE Transactions on Circuits and Systems I: Regular Papers*, *52*(12), pp.2726-2738.

Lopes-Dias, C., Sburlea, A.I. and Müller-Putz, G.R., 2019. Online asynchronous decoding of error-related potentials during the continuous control of a robot. *Scientific reports*, *9*(1), pp.1-9.

Lotte, F., 2015. Signal processing approaches to minimize or suppress calibration time in oscillatory activity-based brain–computer interfaces. *Proceedings of the IEEE*, *103*(6), pp.871-890.

Ma, B.Q., Li, H., Luo, Y. and Lu, B.L., 2019, July. Depersonalized cross-subject vigilance estimation with adversarial domain generalization. In 2019 International Joint Conference on Neural Networks (IJCNN) (pp. 1-8). IEEE.

Marathe, A.R., Ries, A.J., Lawhern, V.J., Lance, B.J., Touryan, J., McDowell, K. and Cecotti, H., 2015. The effect of target and non-target similarity on neural classification performance: a boost from confidence. *Frontiers in neuroscience*, *9*, p.270.

Mei, N., Grossberg, M.D., Ng, K., Navarro, K.T. and Ellmore, T.M., 2018. A high-density scalp EEG dataset acquired during brief naps after a visual working memory task. *Data in brief*, *18*, pp.1513-1519. Available: osf.io/chav7

McDaniel, J.R., Gordon, S.M. and Lawhern, V.J., 2020, October. Modeling the Relationship Between Cognitive State and Task Performance in Passive BCIs using Cross-Dataset Learning. In *2020 IEEE International Conference on Systems, Man, and Cybernetics (SMC)* (pp. 4020-4025). IEEE.

McDaniel, J.R., Gordon, S.M., Solon, A.J. and Lawhern, V.J., 2018, July. Analyzing p300 distractors for target reconstruction. In *2018 40th Annual International Conference of the IEEE Engineering in Medicine and Biology Society (EMBC)* (pp. 2543-2546). IEEE.

McDowell, K., Lin, C.T., Oie, K.S., Jung, T.P., Gordon, S., Whitaker, K.W., Li, S.Y., Lu, S.W. and Hairston, W.D., 2013. Real-world neuroimaging technologies. *Ieee Access*, *1*, pp.131-149.

Metcalfe, J.S., Marathe, A.R., Haynes, B., Paul, V.J., Gremillion, G.M., Drnec, K., Atwater, C., Estepp, J.R., Lukos, J.R., Carter, E.C. and Nothwang, W.D., 2017, May. Building a framework to manage trust in automation. In *Micro-and nanotechnology sensors, systems, and applications IX* (Vol. 10194, p. 101941U). International Society for Optics and Photonics.

Niculescu-Mizil, A. and Caruana, R., 2005, August. Predicting good probabilities with supervised learning. In Proceedings of the 22nd international conference on Machine learning (pp. 625-632).

Parsons, T.D., 2015. Virtual reality for enhanced ecological validity and experimental control in the clinical, affective and social neurosciences. Frontiers in human neuroscience, 9, p.660.

Pfurtscheller, G. and Neuper, C., 2001. Motor imagery and direct brain-computer communication. Proceedings of the IEEE, 89(7), pp.1123-1134.

Philiastides, M.G., Ratcliff, R. and Sajda, P., 2006. Neural representation of task difficulty and decision making during perceptual categorization: a timing diagram. *Journal of Neuroscience*, *26*(35), pp.8965-8975.

Polich, J., 1998. P300 clinical utility and control of variability. *Journal of Clinical Neurophysiology*, *15*(1), pp.14-33.

Ries, A.J., Touryan, J., Ahrens, B. and Connolly, P., 2016. The impact of task demands on fixation-related brain potentials during guided search. *PloS one*, *11*(6), p.e0157260.

Ries, A.J., Touryan, J., Vettel, J., McDowell, K. and Hairston, W.D., 2014. A comparison of electroencephalography signals acquired from conventional and mobile systems. *Journal of Neuroscience and Neuroengineering*, *3*(1), pp.10-20.

Salazar-Gomez, A.F., DelPreto, J., Gil, S., Guenther, F.H. and Rus, D., 2017, May. Correcting robot mistakes in real time using EEG signals. In *2017 IEEE international conference on robotics and automation (ICRA)* (pp. 6570-6577). IEEE.

Sawaki, R. and Katayama, J.I., 2006. Stimulus context determines whether non-target stimuli are processed as task-relevant or distractor information. *Clinical Neurophysiology*, *117*(11), pp.2532-2539.

Shorten, C. and Khoshgoftaar, T.M., 2019. A survey on image data augmentation for deep learning. *Journal of Big Data*, *6*(1), pp.1-48.

Solon, A.J., Lawhern, V.J., Touryan, J., McDaniel, J.R., Ries, A.J. and Gordon, S.M., 2019. Decoding P300 variability using convolutional neural networks. *Frontiers in human neuroscience*, *13*, p.201.

Sonkusare, S., Breakspear, M. and Guo, C., 2019. Naturalistic stimuli in neuroscience: critically acclaimed. Trends in cognitive sciences, 23(8), pp.699-714.

Stock, A.K., Popescu, F., Neuhaus, A.H. and Beste, C., 2016. Single-subject prediction of response inhibition behavior by event-related potentials. *Journal of neurophysiology*, *115*(3), pp.1252-1262.

Su, K. and Robbins, K.A., 2013, August. A framework for content-based retrieval of EEG with applications to neuroscience and beyond. In The 2013 International Joint Conference on Neural Networks (IJCNN) (pp. 1-8). IEEE.

Szafir, D. and Mutlu, B., 2012, May. Pay attention! Designing adaptive agents that monitor and improve user engagement. In *Proceedings of the SIGCHI conference on human factors in computing systems* (pp. 11-20).

Vahid, A., Mückschel, M., Neuhaus, A., Stock, A.K. and Beste, C., 2018. Machine learning provides novel neurophysiological features that predict performance to inhibit automated responses. *Scientific reports*, *8*(1), pp.1-15.

Vahid, A., Mückschel, M., Stober, S., Stock, A.K. and Beste, C., 2020. Applying deep learning to single-trial EEG data provides evidence for complementary theories on action control. *Communications biology*, *3*(1), pp.1-11.

van Atteveldt, N., van Kesteren, M.T., Braams, B. and Krabbendam, L., 2018. Neuroimaging of learning and development: improving ecological validity. Frontline Learning Research, 6(3), p.186.

Watson, T.D., Azizian, A., Berry, S. and Squires, N.K., 2005. Event-related potentials as an index of similarity between words and pictures. *Psychophysiology*, *42*(4), pp.361-368.

Wu, D., Lawhern, V.J., Gordon, S., Lance, B.J. and Lin, C.T., 2016. Driver drowsiness estimation from EEG signals using online weighted adaptation regularization for regression (OwARR). *IEEE Transactions on Fuzzy Systems*, *25*(6), pp.1522-1535.

Wu, D., Chuang, C.H. and Lin, C.T., 2015, September. Online driver's drowsiness estimation using domain adaptation with model fusion. In *2015 International Conference on Affective Computing and Intelligent Interaction (ACII)* (pp. 904-910). IEEE.

Zander, T.O. and Kothe, C., 2011. Towards passive brain–computer interfaces: applying brain–computer interface technology to human–machine systems in general. *Journal of neural engineering*, *8*(2), p.025005.

Zhang, Z., Duan, F., Sole-Casals, J., Dinares-Ferran, J., Cichocki, A., Yang, Z. and Sun, Z., 2019. A novel deep learning approach with data augmentation to classify motor imagery signals. *IEEE Access*, *7*, pp.15945-15954.